\documentclass{sf2a-conf2018}
\usepackage{graphicx}
\usepackage{hyperref}
\usepackage[]{natbib}  
\usepackage{epstopdf}

\def\BibTeX{{\rm B\kern-.05em{\sc i\kern-.025em b}\kern-.08em
    T\kern-.1667em\lower.7ex\hbox{E}\kern-.125emX}}
\bibpunct{(}{)}{;}{a}{}{,}  %%%%%%%%%%%%%  A&A bibliography style
\newcommand{\kms}{\ensuremath{\mathrm{km}~\mathrm{s}^{-1}}}
\newcommand{\teff}{\ensuremath{T_{\rm eff}}}
\newcommand{\logg}{\ensuremath{\log g}}
\newcommand{\vsini}{\ensuremath{v \sin{i}}}

\begin{document}

\TitreGlobal{SF2A 2018}

%%-----------------------------------------------------------------
%%      the top matter
%%

\title{Line identifications in the spectrum of $\chi$ Lupi A}

\runningtitle{Abundances in the atmosphere of $\chi$ Lupi A}

\author{R. Monier}\address{LESIA, UMR 8109, Observatoire de Paris Meudon, Place J.Janssen, Meudon, France}
\author{T. K{\i}l{\i}co{\u g}lu} \address{Ankara University, Faculty of Sciences, Department of Astronomy and Space Sciences, Ankara, Turkey}

%% IF Author3 has the same affiliation than Author1:
%\author{C.\,E. Author3$^1$}

%% IF Author3 has its own affiliation:
%\author{C.\,E. Author3}\address{Dept. of Chess, University of Games, 35101 Las Vegas, Monaco} 

%% IF Author3 has two affiliations, the one of Author1 and a second one:
%\author{C.\,E. Author3$^{1,}$}\address{Dept. of Chess, University of Games, 35101 Las Vegas, Monaco} 

%% Keep this line, even if the page will be settled afterwards.
\setcounter{page}{237}

%%-----------------------------------------------------------------

\maketitle

%%-----------------------------------------------------------------
%%        The abstract
%% 
%%  Warning!  within the abstract:
%%  - do not use macros. 
%%  - do not use commands like: \cite, \citet, \citep ... etc.

\begin{abstract}
We present new abundance determinations for the sharp-lined HgMn star $\chi$ Lupi A from archival FEROS spectra. Selected unblended lines with accurate atomic data have been synthesized to derive new abundances for $\chi$ Lupi A. These spectra show evidence of the presence of blue-shifted lines of the companion $\chi$ Lupi B. The synthesis of the spectrum of $\chi$ Lupi B confirms that this star is cooler, probably an early A star with normal abundances.
\end{abstract}

%% Insert the keywords (to appear in the ADS indexing)
%% Keywords must be separated by a comma

%%-----------------------------------------------------------------

\section{Introduction}
%%---------------------

Previous studies of HgMn star $\chi$ Lupi A have reported overabundances of iron-peak elements in its atmosphere and pronounced overabundances of heavy elements. 
The last extensive abundance analysis from optical spectra is that of \citet{wahlgren1994}.
The low rotational velocity of this star facilitates continuum placement and line synthesis.
It also favors a radiative atmosphere little mixed by rotation. 
Overabundances and underabundances probably reflect an efficient action of radiative acceleration on these heavy elements which have rich transitions, accumulating these elements in the line forming region.

The aim of this work is therefore to provide determinations of new abundances of heavy elements, using upgaded atomic data. As our spectra obviously show the presence of the blue-shifted lines of the companion $\chi$ Lupi B, we have also attempted to model the lines of $\chi$ Lupi B.

\section{Observed spectra and reduction}

The observed FEROS spectrum ($R=48000$) of $\chi$ Lupi has been retrieved from the ESO archive. This FEROS spectrum spans a wide wavelength range from 3700 \AA\ up to 7500 \AA. The exposure time of the spectrum is 50 seconds and the signal-to-noise ratio is 325.

\section{Synthetic spectrum computations and abundance determinations}

The fundamental parameters have been derived using the UVBYBETA program \citep{napiwotzki1993}.
For $\chi$ Lupi A, this yields $\teff=10608 \pm\ 200$ K, $\vsini=5.0\pm0.5\ \kms$, $\logg=3.98 \pm0.25$ dex.
We have derived a microturbulence velocity of $\xi=0.10 \pm0.20\,\kms$ by requesting that strong and weak lines of {\small Fe\,II} yield the same iron abundance.

We computed a model atmosphere with the {\small ATLAS9} code \citep{kurucz1993} with 72 parallel layers assuming Local Thermodynamical Equilibrium (LTE), Radiative Equilibrium (RE) and Hydrostatic Equilibrium (HE). Synthetic spectra were computed using {\small SYNSPEC49/SYNPLOT} \citep{hubenylanz1995} code by using as first solution the solar abundances. In order to compute the composite spectrum consisting of the spectra of A and B components, we modified {\small SYNPLOT} interface for binary stars, into a new interface which we call {\small SYNPLOTBIN}. This interface computes the flux spectrum of the components individualy using {\small SYNSPEC49}, combines them and then normalizes them using the theoretical continuum fluxes for the given atmospheric parameters. 

\section{The derived abundances}

For $\chi$ Lupi A, we find distinct underabundances of He, C, nearly solar abundances for O, Mg, Al,  S, Ca, Sc, and Fe. We find mild overabundances for P and most of the iron-peak elements. We find pronounced overabundances for the the Sr-Y-Zr triad, Ba  and Hg (about 100000 $\odot$).
Using the found abundances, we are preparing the first list of identifications for all lines absorbing more than 2 \% in the spectrum of $\chi$ Lupi A from 3700 \AA\ up to 7500 \AA. The modelling of the the lines of $\chi$ Lupi B suggests this star is a A3 dwarf with a normal surface composition and we confirm the atmospheric parameters of \teff=9200 K and \logg = 4.00 found by \citet{wahlgren1994}.
Figure 1 shows the composite synthetic spectrum for $\chi$ Lupi A and B superimposed onto the observed spectrum. At the time of the observation, $\chi$ Lupi A was redshifted with an orbital radial velocity of $15~\kms$ while $\chi$ Lupi B was blueshifted with a velocity of $-56\ \kms$.

% figure
\begin{figure}[ht!]
 \centering
 \includegraphics[width=0.8\textwidth,clip]{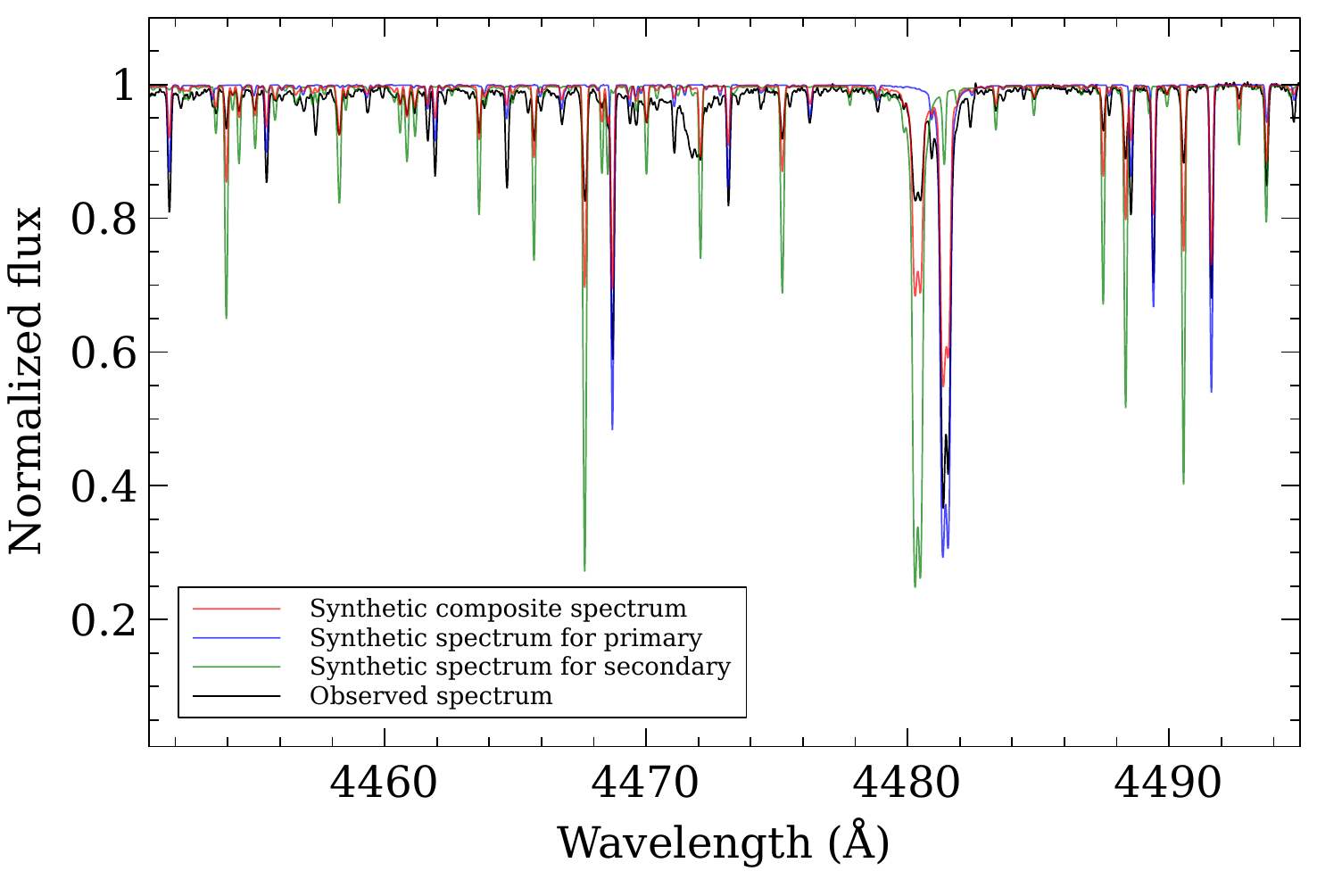}      
%% Note the ABSENCE of the extension .pdf  !
  \caption{Line synthesis of $\chi$ Lupi A and $\chi$ Lupi B in the range 4450-4500 \AA}
  \label{author1:fig1}
\end{figure}

\section{Conclusions}
%%--------------------

Using the upgraded atomic data, we have derived a new set of abundances using a high resolution, high signal-to-noise FEROS spectrum of $\chi$ Lupi A+B. The synthesis of the blue-shifted lines of $\chi$ Lupi B confirms that it is a superficially normal A3 dwarf.
% Optional acknowledgements
% -------------------------
\begin{acknowledgements}
We acknowledge use of the ESO archive facility at http://archive.eso.org
\end{acknowledgements}

\bibliographystyle{aa}  % A&A bibliography style file (aa.bst)
\bibliography{monierkilicoglu} % your references in file: Yourfile.bib

\end{document}